\documentstyle[prd,aps,epsfig,twocolumn]{revtex}

\newcommand{\lab}[1]{\label{#1}}

\newcommand{\fslash}[1]{#1
  \hspace{-5.5pt}\raisebox{-0.2pt}{\slash}\hspace{2pt}}

\begin{document}

\title{Dynamical screening away from equilibrium: \\ hard photon production and
  collisional energy loss }

\author{R.~Baier and  M.~Dirks\thanks{talk presented at 5th International 
 Workshop on Thermal Field Theories and their Applications, Regensburg, 
 Germany, August 1998}}
\address{
Fakult\"{a}t f\"{u}r Physik, Universit\"{a}t Bielefeld,
D-33501 Bielefeld, Germany}

\author{K.~Redlich}
\address{Institute for Theoretical Physics, University of Wroclaw, \\
PL-50204 Wroclaw, Poland\\and\\
GSI, PF 110552,  D-64220 Darmstadt, Germany}

\maketitle

\begin{abstract}
We investigate the production rate for hard real photons and the collisional
energy loss in the quark-gluon plasma away from chemical equilibrium. 
Applying the Hard-Thermal-Loop resummation scheme away 
from equilibrium, we can show that Landau damping provides dynamical screening
for both fermion and boson exchange present in the two quantities.
\end{abstract}

\section{Introduction}
Considerable effort is being made to experimentally 
realize a new 
deconfinement phase of hadronic matter, the quark-gluon
plasma, in ultra-relativistic heavy ion collisions. 
In parallel important progress has also been made  in the theoretical
understanding of this phase which is central for the experimental
detection to be successful. For an equilibrated
system in particular the Hard-Thermal-Loop (HTL) resummation scheme 
\cite{htl1,htl2,htl3} allows 
to take consistently
into account collective effects on the soft scale within (resummed)
perturbation theory \cite{michel:book}. 
A number of important quantities, with unscreened mass singularities 
preventing any prediction
within a fixed order perturbative calculation, can be obtained 
consistently in this scheme. 

It has been realized 
that a valid comparison of theoretical predictions with
experimental data should include non-equilibrium effects 
beyond the simplified assumption of thermal equilibrium, because a
fully equilibrated plasma might not be realized.
In this work we therefore explore the role of collective soft scale phenomena
away from equilibrium and the  generalization of the HTL resummation scheme. 
This will be done on the relevant examples of hard photon production 
\cite{kapusta:photon,baier:photon,shuryak:nephoton,strickland:nephoton,kaempfer:nephoton,thoma:nephoton,muller:nephoton} and the
collisional energy loss \cite{bjorken:engl,braaten:engl1,braaten:engl2} 
which are known to be sensitive to soft scale physics.
The following discussion is based on \cite{baier:nephoton,baier:neegloss}
where further details can be found. 

\section{Approximation scheme: Chemical non-equilibrium}
Our analysis assumes the scenario of chemical non-equilibrium 
that has been found to be a relevant, tractable approximation to the 
pre-equilibrium dynamics in heavy ion collisions
\cite{shuryak:hotglue,biro:therma,kapusta:cascade,wong:therma}. It is
inspired from an analysis of the equilibration process in a heavy ion collision
where the elastic momentum exchange is found to be much more effective than the
inelastic particle production. While local thermal equilibrium might 
therefore be
quickly established the parton densities are predicted to stay away from
equilibrium through to the hadronization of the plasma \cite{biro:therma}. 
Thereby the
undersaturation of quarks will be especially dramatic as has been formulated
in the hot-glue scenario \cite{shuryak:hotglue}. 
Following \cite{biro:therma} and as in 
\cite{thoma:nephoton,kaempfer:nedilepton} we parameterize this scenario 
in terms of factorized fugacities inserted into the distributions which 
are assumed to be equilibrated in momentum space
\begin{eqnarray}
 & &\tilde n(X, p) = \left\{ 
 \begin{array}{l} 
 \lambda_q(X) n_F(|p_0|), \quad p_0 >0 \\
 1- \lambda_q n_F(|p_0|), \quad p_0 < 0
 \end{array}\right. , \nonumber \\
 & &  n(X,p) = \left\{
 \begin{array}{l}
 \lambda_g(X) n_B(|p_0|), \quad p_0 > 0  \\
 -[ 1+\lambda_g(X) n_B(|p_0|)], \quad p_0 < 0.
\end{array} \right.  \lab{eq:nlam}
\end{eqnarray}
We do not attempt to investigate the microscopic non-equilibrium
evolution of the distribution functions in the quark-gluon plasma, the 
evolution of the fugacities $\lambda(X)$ on the large scale $X$ of inelastic
processes is taken as input e.g. from \cite{biro:therma}. On this background we calculate 
the emission and absorbtion rates for probes external to the plasma such as
electromagnetic rates and the collisional energy loss of a projectile.  
For the description of the dynamics we rely on the Closed-Time-Path formulation
of statistical field theory \cite{landsman:rev,chou:noneq}. The simplified scenario of chemical
non-equilibrium allows for the standard approximation scheme of
neglecting any but the second order non-equilibrium correlations and performing
a gradient expansion 
\cite{chou:noneq,heinz:noneq,calzetta:noneq,henning:noneq}, here in 
terms of the large scale $X\sim 1/g^4 T$ 
of the inelastic particle production. Scales fast with respect to this, 
in particular both
the hard $(\sim 1/T)$ and soft $(\sim 1/gT), \, g <  1$,  scales involved in the 
HTL resummation scheme, 
can be transformed to momentum space. To lowest order in the gradients a 
perturbative expansion can be set up very similar to the equilibrium 
case with free in medium propagators depending on the modified distributions
Eq.~(\ref{eq:nlam}). In terms of Wigner variables $(X,p)$: 
\begin{eqnarray}
 & &iD_{21} = 2\pi \varepsilon(p_0) \delta(p^2) (1+n(X,p_0)) ,\nonumber \\
 & &iS_{21}(X,p) = \fslash{p} 2\pi\delta(p^2) \varepsilon(p_0) 
 (1-\tilde n(X,p_0)). \lab{eq:propfree}
\end{eqnarray}
Predictions for absorption and emission rates as calculated in this
approximation scheme will incorporate the dependence on $\lambda(X)$ locally.
As in
the equilibrium situation, finite results for the real hard photon spectrum 
and the collisional energy loss  
can be expected only after resummation of leading HTL corrections.
 Away from equilibrium taking higher order
corrections into account brings about additional terms in the
perturbative expansion \cite{altherr:pinch1,altherr:pinch2} which vanish  
due to detailed balance 
in the equilibrium situation. A careful analysis is therefore required in order
to demonstrate that Landau damping provides dynamical screening of the
singularities also away from equilibrium.

\section{soft fermion exchange in hard photon production}
\begin{figure*}
\begin{center}
(a)
\epsfig{file=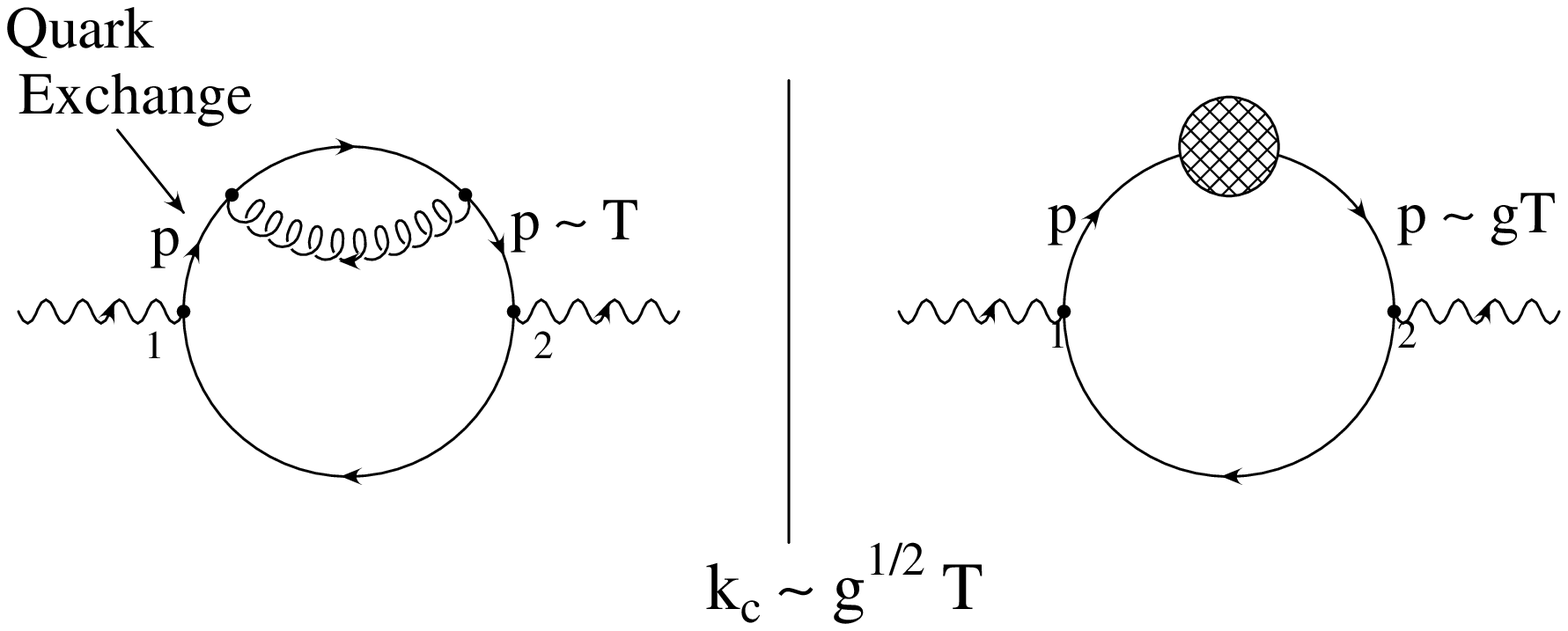, width=7cm}
(b)
\end{center}
\caption{\label{fig:photon}Diagrams contributing to hard real photon production
  in fixed leading order for hard exchanged momentum (a) and in HTL-resummed
 perturbation theory for soft exchanged momentum (b). The parameter $k_c$,
 separating hard and soft scales is introduced in intermediate steps of the
 calculation \protect\cite{braaten:screening}  }
\end{figure*}

\lab{sec:photon}
The production of hard real photons to lowest order arises from 
annihilation and Compton scattering. The rate can be obtained form the
absorptive part of the two-loop approximation to the photon selfenergy 
as shown in Fig. \ref{fig:photon}.
The exchange of a massless quark in these processes induces a logarithmic mass
singularity when calculating to fixed lowest order, Fig. \ref{fig:photon}(a). 
In equilibrium a solution to this problem is found within the HTL resummation
scheme that accounts for the fact that for soft exchanged momentum the dispersion
of the exchanged particle is modified by leading interaction effects. 
Within the resummed theory in particular Landau damping effects are included
that dynamically screen the singularity on the soft scale $m_q \sim gT$
\cite{kapusta:photon,baier:photon}.
Away from equilibrium, in leading order of the gradient expansion, 
the resummed fermion propagator can be written 
\begin{eqnarray}
 S_{12}^\star (p)    &=& - \tilde n(X,p_0)
 (\frac{1}{\fslash p -\Sigma_R + i\varepsilon p_0} -
        \frac{1}{\fslash p -\Sigma_R^\star - i\varepsilon p_0})  \nonumber \\
 &-&  \frac{1}{\fslash p - \Sigma_R + i\varepsilon p_0 }
  \left[ (1-\tilde n(X,p_0)) \Sigma_{12} \right. \nonumber \\
 & & \qquad \left. + \tilde n(X,p_0) \Sigma_{21} 
  \right] \frac{1}{\fslash p - \Sigma_R^\star -i\varepsilon p_0} , \lab{eq:propf}
\end{eqnarray}
with a first term having the structure of the equilibrium expression and a
second term being nonzero only when detailed balance is violated,  
$\Sigma_{12} \neq -  e^{-p_0/T} \Sigma_{21}$. 
Two remarks have to be added about the modified structure of the propagator,
Eq.~(\ref{eq:propf}).
First we find that the non-equilibrium modification of the dispersion 
as defined by the pole of the retarded propagator in the first term of
Eq.~(\ref{eq:propf}) can be absorbed into a redefinition of the soft scale
parameter used in the equilibrium situation
\begin{eqnarray}
  \tilde m_q^2 &=& \frac{g^2}{2\pi^2} C_F \int_0^\infty E dE (n(X,E) + \tilde n(X,E))
  \nonumber \\
 &=& \frac{g^2 T^2}{12} C_F \left(\lambda_g + \frac{\lambda_q}{2}\right) .
 \lab{eq:newmq}
\end{eqnarray}
This arises, because, as in the equilibrium analysis, the leading
HTL contribution to the one loop selfenergy is found to be
proportional to the radial integral used in the definition Eq.~(\ref{eq:newmq})
that contains the only $\lambda$-dependence involved. 
Secondly, in order to discuss the role of the extra contribution to the
propagator Eq.~(\ref{eq:propf}) it is convenient to decompose the expression
in square brackets 
\begin{eqnarray}
& &(1-\tilde n(X,p_0)) \Sigma_{12} + \tilde n(X,p_0) \Sigma_{21} \nonumber\\
& & \qquad  = 
  (1-2\tilde n(X,p_0)) \Sigma^- + \Sigma^+
\end{eqnarray}
into a contribution from $\Sigma^- = \frac{1}{2}(\Sigma_{12} - \Sigma_{21} ) =
iIm \Sigma_R$ plus a
remainder $\Sigma^+ = \frac{1}{2} \left(\Sigma_{12} + \Sigma_{21}\right)$.
In the fermionic case the contribution from $\Sigma^+$ is found to be 
suppressed with respect to the dominant HTL scale $g^2T^2$ as set by
$\Sigma^-$, Eq.~(\ref{eq:newmq}). From an explicit calculation 
\begin{eqnarray}
& &\Sigma^+ (X,p) \sim p_0 R\, \Sigma^-(X,p) \ll \Sigma^- , \nonumber \\
& & R = \frac{ \int_0^\infty dk~  k~  \frac{\partial}{\partial k} 
  \tilde n(k) (1+2n( k)) }{ \int_0^\infty dk~  k~ [n(k) + \tilde n(k)]} 
 \sim \frac{1}{T} ,
\end{eqnarray}
for soft $p_0 \sim gT$. 
Within the HTL resummation scheme $\Sigma^+$ can therefore be neglected in
the resummed propagator Eq.~(\ref{eq:propf}) and  with 
\begin{equation}\lab{eq:rearf}
 2\Sigma^- = \Sigma_R -\Sigma_A = 
  {S_A^\star}^{-1} - {S_R^\star}^{-1} 
\end{equation}
the remaining terms are rearranged into
\begin{eqnarray}
\left.  S_{12}^\star \right|_{HTL} &\simeq& -\tilde n(X,p_0) \left(  
 S_R^\star -  S_A^\star\right) \nonumber \\
 & &\quad   - (\frac{1}{2}-\tilde n(X,p_0)) S_R^\star [2\Sigma^-] S_A^\star
 \nonumber \\
 &=& - \frac{1}{2} \left(S_R^\star - 
 S_A^\star \right).
\lab{eq:prosoft}
\end{eqnarray}
Given the close analogy of this effective resummed propagator with the
equilibrium expression, the remaining steps are performed in parallel with the
equilibrium calculation \cite{kapusta:photon,baier:photon}
 taking only the modified dispersion into account. 
We find the rate for hard photon production to depend on the modified
scale parameter Eq.~(\ref{eq:newmq}) that enters as a prefactor and 
screens the logarithmic singularity: 
\begin{eqnarray}
  E_\gamma \frac{dR}{d^3q} &=& e_q^2 \frac{\alpha\alpha_s}{2\pi^2}
 {\lambda_q} T^2
   e^{-E_\gamma/T} \nonumber \\
  & & \quad \, \times\left[ \frac{2}{3}{(\lambda_g + \frac{\lambda_q}{2})}
  \ln \left(\frac{2E_\gamma T}{{\tilde m_q^2(\lambda_q,\lambda_g)}} 
\right)\right.
 \nonumber \\ 
 & & \left. \qquad + \frac{4}{\pi^2} C(E_\gamma ,
  T, \lambda_q, \lambda_g) \right] ,  \lab{eq:resph}
\end{eqnarray}
where  the nonsingular contribution $C(E_\gamma ,
  T, \lambda_q, \lambda_g)$ has been calculated explicitly in
\cite{baier:nephoton}. 

The result Eq.~(\ref{eq:resph}) is independent of the 
separating scale $k_c$ introduced in
intermediate steps of the calculation by matching the soft
partial result with the result from a fixed order calculation for hard
exchanged momentum. In this context it is important to point out that no
additional singularities are showing up away from equilibrium although at 
one loop order 
the extra term  in the fermion 
propagator analogous to Eq.~(\ref{eq:propf}) is pinch singular on the mass
shell \cite{altherr:pinch1}.
 The important observation is that for hard real photon production the 
singular configuration lies only on the boundary of the available phase space
that is restricted to spacelike momentum exchange. The effective propagator
for the hard quark exchange to
be used in the upper part of Fig. \ref{fig:photon}(a) follows from
Eq.~(\ref{eq:propf}) in the one loop approximation,
 
\begin{equation}
\left. \delta S_{12}(p) \right|_{p^2 \le -k_c^2} 
   \stackrel{\wedge}{=}
   -\frac{1}{(p^2)^2} \fslash p \Sigma_{12} \fslash p .
\end{equation}
There remains no explicit dependence on the distribution $\tilde n (X,p_0)$.  A mass 
singularity is induced, which, however,  is dynamically screened upon matching with
the contribution from the soft exchange with the propagator Eq.~(\ref{eq:prosoft}).

\section{Soft (gauge) boson exchange -- collisional energy loss}
\begin{figure}
\begin{center}
(a)
\epsfig{file=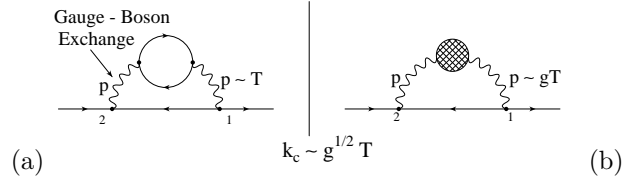, width=7cm}
(b)
\end{center}
\caption{\label{fig:egloss}Diagrams contributing to the collisional energy loss
  in fixed leading order for hard exchanged momentum (a) and in HTL-resummed
 perturbation theory for soft exchanged momentum (b). } 
\end{figure}

The collisional energy loss of a propagating heavy  fermion with mass $M$
\cite{braaten:engl1,braaten:engl2}, which is not thermalized,  is obtained 
from the absorptive
part of the fermion selfenergy,
i.e. from its $21$-component \cite{michel:book}. The corresponding  diagrams
are  shown in Fig.~\ref{fig:egloss}. 
To lowest order the loss arises from the elastic interaction with the medium
constituents via the exchange of one gauge boson, Fig.~\ref{fig:egloss}(a);
again mass singularities are known to arise for 
massless photon/gluon exchange.

It is important to recall that the energy loss is obtained form these diagrams
only upon weighting the loop integration with one extra power of the energy 
exchanged:
\begin{eqnarray}
  - \frac{dE}{dx} &=& - \frac{1}{4E} Tr\left[ (\fslash q + M) 
 i\Sigma'_{21}(X,q)\right] 
 \nonumber \\
 &=& \frac{e^2}{4E} \int \frac{d^3 \vec{p}}{(2\pi)^4}
   \int\frac{dp_0 p_0}{v} \nonumber \\
 & & \quad \times Tr[(\fslash q + M) \gamma_\mu iS_{12}^{T=0} (p-q) 
 \gamma_\nu]
 iD_{21}^{\mu\nu}(X,p).
\lab{eq:egl1}
\end{eqnarray}
The extra power of $p_0/v$ distinguishes $dE/dx$ from the interaction/damping
rate and makes it well defined in equilibrium within the resummed theory.

Away from equilibrium the resummed propagator for soft gauge boson-exchange
has a modified structure similar to the fermion propagator
\cite{thoma:noneq1}; 
longitudinal and transverse components in Coulomb gauge read
\begin{eqnarray*}
 D_{21}^{\star \scriptscriptstyle{L/T}} &=& (1+n(X,p_0)) 
 (D_R^{\star \scriptscriptstyle{L/T}} - D_A^{\star \scriptscriptstyle{L/T}} )
  \nonumber \\
 & & \hspace*{-0.8cm} + D_R^{\star \scriptscriptstyle{L/T}} \left[ n(X,p_0) 
 \Pi_{21}^{\scriptscriptstyle{L/T}} - (1+n(X,p_0)) \Pi_{12}^{\scriptscriptstyle{L/T}}\right]
 D_A^{\star \scriptscriptstyle{L/T}}.
\end{eqnarray*}
The discussion follows the lines of Sec. \ref{sec:photon}. The modifications
of the dispersion relation in HTL approximation can again be absorbed
into a redefinition of the plasma frequency according to 
\begin{equation}\lab{eq:newmb}
m_\gamma^2 = \frac{e^2 T^2}{9} \quad\to\quad
 \tilde m_\gamma^2 = \frac{4e^2}{3\pi^2} \int_0^\infty dk k \tilde n(X,k) = 
  \lambda_f m_\gamma^2 
\end{equation}
in QED.
For the role of the extra term however an important difference with the 
fermionic case arises. In the bosonic case the contribution from 
$\Pi^+= \frac{1}{2}(\Pi_{12} + \Pi_{21})$ is found to dominate over 
$\Pi^- = \frac{1}{2}(\Pi_{12} - \Pi_{21}) = iIm \Pi_R$ for both longitudinal
and transverse components \cite{thoma:noneq1}
\begin{eqnarray}
& &-\Pi^+_{\scriptscriptstyle{L/T}} = \frac{1}{p_0}~ R~
\Pi^-_{\scriptscriptstyle{L/T}} ,  \nonumber \\
& &R = \frac{\int_0^\infty dk k^2 \tilde n(k) (1-\tilde n(k))}
 {\int_0^\infty dk k\tilde n(k)} \sim T , 
\end{eqnarray}
and therefore cannot be neglected in general. 
However, both contributions have a symmetry under $p_0 \to -p_0$,
while $\Pi^-$ is odd, $\Pi^+$ is even. Given 
the additional power of $p_0$ present in the energy loss rate the 
contribution of $\Pi^+$ therefore integrates to zero in this special case as
\begin{equation}
\int_{-vp}^{vp} dp_0 { p_0} \Pi^+(X,p_0) = 0 .
\end{equation}
As a result the dominant contribution to $dE/dx$ comes from $\Pi^-$ and the
corresponding terms in the resummed propagator rearrange into 
\begin{equation}
\left. D_{21}^{\star \scriptscriptstyle{L/T}} ~\right|_{dE/dx} 
 \stackrel{\wedge}{=} \frac{1}{2}
   (D_R^{\star \scriptscriptstyle{L/T}} - D_A^{\star \scriptscriptstyle{L/T}})
 \qquad \mbox{in} \quad dE/dx, 
\end{equation}
which is similar in structure to the equilibrium expression and allows to
perform the remaining steps as in \cite{braaten:engl1}. 
As for hard photon production
no additional (pinch) singularities arise away from equilibrium for spacelike
momentum exchange. For energies $E \ll M^2/T$
we find the energy loss away from equilibrium to be well defined 
\begin{eqnarray}\lab{eq:reseg}
-\frac{dE}{dx} &=& \frac{ 3 e^2 \tilde m_\gamma^2}{8\pi v} 
  \left(1-\frac{1-v^2}{2v}
  \ln \frac{1+v}{1-v}\right) 
 \left[ \ln \frac{ET}{3M{  \tilde m_\gamma}}\right. \nonumber \\
 & &  \hspace*{2cm}+ A(v)  \Bigr],
\end{eqnarray}
with Landau damping dynamically screening the logarithmic mass singularity. 
Following the calculation for the equilibrium case 
it turns out that the regular contribution $A(v)$ 
does not depend on the fugacity factor $\lambda_f$, and that it is therefore the same as discussed 
in \cite{braaten:engl1}.

The generalization for a heavy quark propagating through a QCD-plasma
 is obtained \cite{baier:neegloss} with the help of the expressions given 
in \cite{braaten:engl2},
\begin{eqnarray}\lab{eq:resegqcd}
-\frac{dE}{dx} &=& \frac{ g^2 \tilde m_g^2}{2\pi v} 
  \left( 1-\frac{1-v^2}{2v}  \ln \frac{1+v}{1-v}\right) 
 \left[ \ln \frac{ET}{3M{  \tilde m_g}}\right. \nonumber \\
 & &  - \frac{ \ln 2}{(1 + \lambda_q N_f / 6 \lambda_g)} + A(v)  \Bigr],
\end{eqnarray}
introducing the QCD coupling $g$ and the thermal gluon mass, including the
proper quark and gluon fugacities:
$\tilde m_g^2 = g^2 T^2 (\lambda_g + \lambda_q \frac{N_f}{6})/3$. 
Because in the QCD case contributions from the scatterings off quarks and gluons
in the plasma have to be added, a dependence on
$\lambda_q$ and $\lambda_g$
is found also in the regular part of Eq.~(\ref{eq:resegqcd}).

\section{Discussion and conclusions}
From the results given in Eqs.~(\ref{eq:resph}) and  (\ref{eq:resegqcd}),
 respectively,
both the hard photon production rate and the collisional energy loss are found
to be suppressed by powers of the fugacity in an undersaturated plasma that is
likely to be created in a relativistic heavy ion experiment
\cite{shuryak:hotglue,biro:therma}. Taking into account
the extra off-equilibrium term in the propagator there is no dependence in the
final result on the distribution of the exchanged particle that would 
have introduced additional powers of fugacities. In the case of the 
photon production  rate this result is especially important phenomenologically
as higher powers of the small quark fugacity could 
seriously question its observability.
It is encouraging that both rates can be obtained consistently in perturbation
theory away from equilibrium with dynamical screening occurring also for
$\lambda\neq 1$. We argued that no additional pinch singularities
\cite{altherr:pinch1}  do
arise in the quantities considered that involve spacelike momentum exchange
only. The treatment given here however is not applicable to 
quantities involving on-shell momentum exchange as e.g. virtual photon
production \cite{baier:pirho,lebellac:pinch,niegawa:dilepton}. 
Further progress in these cases requires to take consistently into account 
the explicit time variations in the distributions
\cite{bedaque:pinch,greiner:pinch} via the inclusion of 
gradient terms which could be neglected in the presented cases.

\section{Acknowledgements}
We like to thank D.~Schiff and M.H.~Thoma
 for helpfull discussions and critical comments. M.D. is
supported by DFG, Contract BA 915/4-2.

\bibliographystyle{prsty}
\bibliography{pro}
\end{document}